\documentclass[a4paper,11pt]{article}

\usepackage{jcappub} 

\usepackage[T1]{fontenc} 
\usepackage{epsf,graphics,subfig,ulem,epsf, amsmath}
\usepackage{amsmath,amssymb,natbib}
\usepackage{lmodern}
\usepackage{cite}

\def\araa{ARA\&A}             
\def\apj{ApJ}                 
\def\apjl{ApJ}                
\def\apjs{ApJS}               
\def\aap{A\&A}                
\def\aapr{A\&A~Rev.}          
\def\mnras{MNRAS}             
\def\pasj{PASJ}               
\def\nat{Nature}              
\def\physrep{Phys.~Rep.}      
\def\ssr{Space~Sc.~Rev.}

\title{\boldmath Survival of dust in super-dusty galaxies at redshifts $z \approx 5-8$}

\author[a1]{Yu. A. Shchekinov,\note{Corresponding author.}}
\author[a]{Biman B. Nath}

\affiliation[a]{Raman Research Institute, Sadashiva Nagar, Bengaluru 560080, India}

\emailAdd{yuri.and.s@gmail.com}

\abstract{The presence of dust in galaxies at redshifts $z>5$ is commonly connected with core collapse supernovae (SN). Galaxies with exceptionally large dust mass, of order of $1 - 3$\% of the galaxy stellar mass, have been detected during the last decade. The required SN dust yield is $\gtrsim 1~M_\odot$ per supernova, which is comparable to the theoretically predicted maximum. However, the reverse shock (RS) penetrating  the SN ejecta significantly destroy the dust particles nucleating there through sputtering. The resulting net dust mass injected into the interstellar gas after processing by the RS turns out to be  $\lesssim 0.1~M_\odot$ per SN. This makes the explanation of the existence of $z>5$ galaxies with dust masses as high as $M_d\gtrsim (0.01-0.03)M_\ast$ a challenging one. In this paper we present arguments in favor of an efficient inhibition of the sputtering  behind the RS, caused by a strong radiation cooling from the dust in the supernova ejecta. }

\begin{document}
\maketitle
\flushbottom

\def\sc{\scriptscriptstyle}
\def\h{H~I}
\def\he{He~I}
\def\hei{He \,II}
\def\12{{1\over 2}}
\def\msun{M_{\odot}}
\def\lsun{L_{\odot}}
\def\div{\nabla\cdot}
\def\grad{\nabla}
\def\rot{\nabla\times}
\def\eg{{\it e.g.,~}}
\def\ie{{\it i.e.,~}}
\def\etal{{\it et~al.,~}}
\def\de{\partial}
\def\msun{M_\odot}
\def\zsun{Z_\odot}
\def\be{\begin{equation}}
\def\ee{\end{equation}}
\def\ltsima{$\; \buildrel < \over \sim \;$}
\def\simlt{\lower.5ex\hbox{\ltsima}}
\def\gtsima{$\; \buildrel > \over \sim \;$}
\def\simgt{\lower.5ex\hbox{\gtsima}}
\def\etal{{et al.\ }} 
\def\red{\textcolor{red}}
\def\blue{\textcolor{blue}}
\def\zsun{Z_\odot} 

\section{Introduction}
\label{intro}
Dust has recently been detected in vast amounts in the early epochs when the Universe was younger than 1 Gyr \citep[][]{Carniani2024}. 
Asymptotic giant branch (AGB) stars, traditionally treated as the dominant source of dust, are unlikely to produce such a large amount of dust within this short time scale. \citet{Rowlands2014} have demonstrated that {dust} production by AGBs stars is sensitive to the initial mass function (IMF) of low and intermediate mass stars, and is at least two orders of magnitude lower than {that} required to explain the dust-to-stellar mass ratio $M_d/M_\ast\equiv\zeta_\ast\sim 10^{-2}$ observed in submillimeter galaxies (SMG) at redshifts $z\approx 1-5$. Analysis of the dust depletion of elements in the ISM of a set of QSO and GRB DLAs at $0.6<z<6.3$  led \citet[][]{Konstantopoulou2024} to conclude that the stellar dust yield can be as small as $\sim 0.01$ of the metal yield in the ISM of galaxies with a metallicity [Z]$<-1.0$. 

The explanation for the `dust budget crisis' {requires} supernovae (SN) as an alternate source of dust. This issue {was} first discussed by \citet[][]{Todini2001,Dunne2003,Morgan2003,Maiolino2004,Dwek2007}, and confirmed further in a series of observations \citep[][ see also  reviews in \citep{Gall2011,Micelotta2018,Schneider2024}]{Michalowski2010,Gall2011a,Valiante2011,Rowlands2014,Peroux2023}. In brief, SNe explosions have been identified as being predominantly responsible for dust production at redshifts higher than $z\sim 5$. However, in several cases,  galaxies at pre-reionization epochs {have been} found to {have} {a} dust-to-stellar mass ratio as high as $\zeta_\ast>0.01$, {so much so} that the observed star formation rate {are not} sufficient for producing it. A growing {number} of galaxies with such an excessive amount of dust have been recently described in \citet[][]{Watson2015,Laporte2017,Hashimoto2019,Hashimoto2023,Bakx2021,Fudamoto2021,Akins2023,Glazer2023,Killi2023,Palla2024}. This tendency of the dust-to-stellar ratio to reach $\simgt 0.01-0.02$ has been recently found in many galaxies in the redshift range $z\simgt 1.5-7$ with an increasing trend towards higher $z$, enhanced by factor of $\sim 3$ in starburst galaxies \citep{Donevski2020,Liao2024,Polletta2024}. These results {essentially} demonstrate that more than a few percents of stellar mass is transformed into dust particles, {which is} challenging to understand.

However, even though SNe explosions {appear to be} the most likely, if not the only, source of dust at these early epochs, this hypothesis {has recently} faced troubles. It has been pointed out that dust particles produced in the SNe ejecta  rapidly {get} destroyed once the ejecta {come in contact with} the surrounding ISM gas \citep{Dwek2007,Nozawa2007,Nath2008,Kirchschlager2019}. Only a tiny fraction of {the} produced dust survives, about $0.01~\msun$, which is far from being sufficient to solve the `dust budget crisis'. 

Observations of young core-collapse supernovae (CCSN) in the local universe {mostly} favour a rather low dust yield: from $M_d\simeq 10^{-4}-10^{-3}~\msun$ \citep{Gall2011,Szalai2013,Tinyanont2019} for SNe that are a few hundred days old,  to $M_d<0.01~\msun$ \citep[][]{Rho2024} for older SNe. {Only a few SNe have been observed to contain larger dust masses than this: (i) SN 1987A with the dust mass $M_d\simeq 0.45-0.8~\msun$ \citep{Dwek2015,Matsuura2015,Wesson2015}; (ii) SN Cas A, with a preliminary estimate $M_d\sim 1~\msun$ \citep{Dunne2009},  and recently revised to $M_d\sim 0.05-0.5~\msun$ \citep{Delooze2017,Priestley2022}, depending of the initial size of injected dust grains, their chemical composition and the fraction of dust destroyed behind the reverse shock; (iii)  Crab SN, with $M_d\sim 0.1-0.25~\msun$ \citep{Bocchio2016} and (iv) $M_d\sim 0.9~\msun$ in the N49 \citep{Bocchio2016}.
}

{Of these four remnants only 1987A seems to represent the pristine ejecta not being processed by the reverse shock \citep{Bocchio2016}.
It is conceivable that the estimated $M_d\simeq 0.45-0.8~\msun$ in the SN 1987A is going to be destroyed after passing the reverse shock into the ejecta. The exact fraction of the destroyed dust will depend on the intrinsic physical and morphological parameters of the SH 1987A. Keeping in mind the fact that the SN 1987A lies nearly in the `main sequence' of the `M$_{dust}$--Age' relation \citep[see Fig. 23 in ][]{Niculescu2022}, the remnant has apparently reached the asymptotic value for the SNe of $t>300$ yr old. If , as concluded by \citet{Bocchio2016}, only  $\sim 0.01-0.08$ dust mass fraction survives the reverse shock attack, then 1987A can supply into the ISM the net dust mass $M_d\simlt 0.7~\msun$. The estimates of the dust survival fraction are sensitive to physical characteristic in the ejecta and in the circumstellar medium \citep{Bianchi2007}, and are rather uncertain.    } 

{In this regard it is worth to note  that recent JWST high-resolution observations of the Cas A remnant indicate that a considerable fraction of dust mass can be locked in the unshocked parts of the ejecta \citep{Milisavljevic2024}. This implies that the currently measured dust mass is yet to be destroyed when the reverse shock will further penerate the filaments. } 

The conflict between the theoretical constraints of dust production by supernovae and the observational requirements in high-redshift galaxies heightens the `dust budget crisis'. It as been clearly demonstrated in recent observational analysis of the required dust mass yield per SN ($y_{sn}$)  in a set of 14 galaxies at $z\simeq 7$ is $0.1\leq y_{sn}\leq 3.3~\msun$. Among these, $\sim 70\%$ have $0.1\leq y_{sn}<0.22~\msun$, and the remaining 30\% have $y_{sn}\geq 1.9~\msun$, as has been inferred by \citet[][ see their Table 2]{Ferrara2022}. One may tentatively conclude that at least around 30\% of galaxies at initial stages of reionization are super-dusty in the sense that their current stellar populations and the related SN rates are not sufficient for providing the necessary amount of dust. 

In this paper, we describe a plausible scenario for the survival of ejected dust particles against unfavourable conditions arising from the interaction of ejecta and the ambient ISM. For this purpose, we consider in detail the thermal properties of the dusty gas in the ejecta at very initial stages of its interaction with the ISM. We show that the destructive effects can be inhibited by fast radiative cooling within the compressed ejecta in the presence of thermal instability. 

In Sec. \ref{prob} we describe the observational background  and explicitly state the puzzling scenario. In Sec. \ref{solut} we discuss possible solutions of the problem which have been recently suggested in the literature. We then discuss our scenario connected with an enhanced dust cooling inside clumps that grow through thermal instability, so that dust particles confined within these dense cold clumps  are shielded  against sputtering. Sec. \ref{summ} summarizes the results.

\section{Where dust in $z>5$ galaxies comes from}\label{prob} 

As mentioned in Introduction, one of the most difficult problems in enhancing the dust yield from SNe is the reverse shock (RS) entering the ejecta \citep{Dwek2007,Nozawa2007,Nath2008}. The RS heats the ejecta layers  up to $\sim 10^8$ K and initiates dust grain disintegration through sputtering, severely diminishing their net yield \footnote{By the net yield is undertood the SN dust yield after processing it by the RS.}. We describe this problem in detail below.

\subsection{Dust destruction by the RS}\label{snfac} 
Consider the galaxy A1689-zD1 ($z=7.13$) as a prototype of a super-dusty galaxy in the pre-reionization epoch. \citet{Bakx2021} estimated the dust mass in A1689-zD1 to be  $M_d\simeq 2\times 10^7~\msun$, with the help of an accurate measurement of the dust temperature and the spectral index $T_d\simeq 40$ K, $\beta_d\simeq 1.6$, making use of the {\it ALMA} Band 9 continuum observations, combined with measurements of the [CII] 158 $\mu$m line emission with an underlying continuum \citep{Sommovigo2020}. {The stellar mass in this galaxy is estimated  \citep{Bakx2020,Bakx2021} to be $M_\ast\sim 3\times 10^9~\msun$, which implies a dust-to-stellar mass ratio $\zeta_\ast\sim 10^{-2}$ and places A1689-zD1 in a sub-population of galaxies at $z\simgt 7$ with an overabundance of dust.} 

Assuming that each SN produces $y_{sn}$ dust mass, the total number of supernovae necessary for production of such a large amount of dust mass is $N_{sn}\sim 2\times 10^7y_{sn}^{-1}$. {On the other hand, the current star formation rate (SFR) in A1689-zD1 is $\dot M_\ast\simeq 10~\msun$ yr$^{-1}$, as inferred from IR and UV \citep{Watson2015}, can produce in 250 Myr (since $z\sim 10$) a total number of SNe as given by $N_{sn}\sim \dot M_\ast\times 250\times 10^6\nu_{sn}\simeq 1.75\times 10^7$, where the specific per mass SN rate $\nu_{sn}\sim 0.007~\msun^{-1}$ is assumed to be \citep[see, e.g., ][]{Madau2001}. The total mass of dust from these SNe $M_d\simeq 2\times 10^7~\msun$ is close to the observed dust content, provided that $y_{sn}\sim 1~\msun$. In other words, the estimates for the total number of SNe produced in the course of star formation and the produced dust mass are consistent {\it only if} $y_{sn}\simgt 1~\msun$.       
}

The dust yield per SN predicted by theoretical studies lies in the range $y_{sn}\sim 0.03-0.6~\msun$ \citep{Todini2001,Bianchi2007,Sarangi2013,Sarangi2015,Lazzati2016,Dwek2019,Marassi2019,Brooker2022}, depending on the SN progenitor mass and assumptions of the nucleation kinetics \citep[see more discussion in recent reviews ][]{Micelotta2018,Schneider2024}. {However, as mentioned above, the net dust yield is a product not only of nucleation in the end of the free-expansion ejecta phase, but also of destruction behind the RS that propagates inward through the ejecta. The temperature behind the RS is $T_0\sim 10^8$ K \citep{Mckee1995,Truelove1999},}
{which implies efficient dust sputtering, at time scales comparable to or shorter than the dynamical time of the ejecta, even for large grains.} As a result, the RS severely destroys recently nucleated dust particles \citep{Dwek2007,Nozawa2007,Nath2008}. The fraction of destroyed dust $f_{d,dest}$ depends on the progenitor mass and the ambient gas density $n_0$. For $n_0\geq 1$ cm$^{-3}$, this fraction can vary from $f_{d,dest}\sim 0.7$ to $\simeq 1$ \citep{Nozawa2007,Nath2008,Hirashita2014} (for more recent discussion see the review by  \citet{Micelotta2018}). 

Later studies based on 2D hydrodynamical simulations, \citet{Slavin2020} have concluded that only $10-20$ \% of silicates and $30-50$ \% carbonaceous grains of large sizes $0.25-0.5~\mu$m ejected by a SN can survive the impact of the reverse shock. A subsequent study with 3D multi-fluid simulations has shown that the result of destruction can be stronger than this: at $n_0\simgt 1$ cm$^{-3}$ the fraction of survived particles with radii $a\simgt 0.3~\mu$m is $f_{d,surv}\simlt 0.3$, and only a fraction  $0.07$ of smaller particles can survive \citep{Kirchschlager2022,Vasiliev2023}. At higher densities the fraction of dust mass that survives  scales as $f_{d,surv}\simeq 0.01y_{\rm sn,_0}n^{-p}$, $p>1$, where $n$ is the ambient density in cm$^{-3}$ \citep{Vasiliev2023}. These numerical estimates put fairly strong upper limits on the net SN dust yield, and make the detection of excessive dust-to-stellar mass ratio difficult to explain. 

In brief, the processing of dust behind the RS critically diminishes the net dust yield from SNe, particularly in growing galaxies at $z>7$ with dense ISM. In such conditions, the value $y_{\rm sn}\sim 0.1~\msun$ would be a too optimistic for an estimate of dust SNe yield. More conservative estimates for the net yield with contributions from SNe and AGB turn out to be  below that of $y_{\rm sn+agb}\simeq 10^{-2}~\msun$ \citep[][]{Valiante2009,Michalowski2010,Hirashita2014,Rowlands2014,Witstok2023}. Taking this effect into account, the expected dust-to-stellar mass ratio in $z>7$ galaxies is considerably reduced:
\be 
\zeta_\ast={M_d\over M_\ast}\sim \nu_{sn}y_{sn}\simlt 0.007\times 10^{-2} \approx 7\times 10^{-5}\,,
\ee
where we have assumed that $M_\ast\sim\dot M_\ast \Delta t$ and $M_d\sim \nu_{\rm sn}y_{\rm sn}\dot M_\ast\Delta t$, $\Delta t$ being the duration of star formation.

\subsection{Alternative sources of dust}\label{alter}

\subsubsection{Dust accretion in the ISM}

It appears that the replenishment of a deficient dust mass in to {\it the diffuse} ISM through its growth in dense clouds as suggested by \citet{Popping2017,Dacunha2023,Narayanan2024} can be an alternative source. However, there are a few issues that make this option debatable. 

The first point is connected with the existence of galaxies that show the value of  $\zeta_\ast$ approaching, or in some cases exceeding, the standard stellar metal yield $\tilde y_Z\sim 10^{-2}$ per stellar mass  \citep{Riechers2013,Wang2019,Donevski2020,Riechers2020,Riechers2013,Marrone2018,Witstok2023}. However, it is straightforward to see that, if we write $f_{\rm ism}= M_g/M_\ast$, then  the  metallicity of the host ISM is nothing but 
\begin{equation}
Z={{\tilde y_Z \times M_\ast -M_d} \over M_g}={ (\tilde y_Z-\zeta_\ast) \over f_{\rm ism}}\,. 
\end{equation}
Clearly, the stellar metal yield $\tilde y_Z$ is the upper limit for  $\zeta_\ast$. 

{There are also problems with the idea of formation of dust through accretion in dense interstellar clouds at the kinetic level. Particles grown in low temperature conditions  become predominantly covered with icy water. This process heavily weakens the sticking ability of other elements on their surface and inhibits further growth \citep{Ferrara2016,Ceccarelli2018}. Moreover, the water mantles are volatile and easily evaporate in the presence of optical and UV photons, which may arise from massive stars either embedded inside the ISM cloud or located outside it. Therefore, the  icy mantles of particles rapidly sublimate when the cloud disintegrates during the process of energetic feedback in ambient diffuse gas. The characteristic sublimation time scale vary in the range $t\sim 10-1000$ yr \citep[][]{Ferrara2016,Ceccarelli2018}.} 

Another obstacle for  dust grains to accrete in the ISM is the critical metallicity needed for dust particles to grow in a gas phase.  The accretion rate on to dust depends on gas metallicity and the star formation rate. In agreement with theoretical predictions, the metallicity threshold is observationally inferred to be  $Z\sim 0.05Z_\odot$, from the dust-to-metallicity correlation in nearby galaxies \citep{Zhukovska2008,Remyruyer2014,Galliano2018,Clark2023,Konstantopoulou2024,Popping2023}. Below this threshold the dust accretion drops by factor of $10\hbox{--}30$, depending on the SFR \citep[][]{Asano2013,Choban2024a,Choban2024b}.

\subsubsection{{Contribution from stars of previous generations}}   

One recently discussed possible alternative is the production of dust during the preceding stellar generations, before the currently observed  phase in $z>7$ galaxies \citet[][]{Lesniewska2019,Tamura2019,Tacchella2022,Witstok2023}. 
For a rough estimate of this hypothesis, one can assume that the metallicity 
in $z\sim 3-7$ galaxies of stellar mass $M_\ast\sim 10^6-3\times 10^9~\msun$ follows the relation,
\be \label{zmrel}
{Z\over Z_\odot}\sim 0.15\left({M_\ast\over 10^9~\msun}\right)^{0.4}\,,
\ee
as can be inferred from \citet[][ see their Fig. 1]{Chemerynska2024}. Here the stellar mass is normalized to $10^9~\msun$, as is typical for $z>10$ galaxies. In this case the mass fraction of metals transformed into dust can be estimated from the fact that the dust-to-metal mass ratio is  $\sim 0.1-0.4$ in the gas metallicity range of $Z/Z_\odot\sim 3\times 10^{-3}-0.3$ \citep[ e.g., Fig. 2 in ][]{Konstantopoulou2024}\footnote{In both cases \citep{Chemerynska2024,Konstantopoulou2024} $Z$ represents the ISM metallicity.}. The interrelations between the metallicity, dust-to-metals and dust-to-gas ratios over a wide range of metallicities $Z/Z_\odot=3\times 10^{-2} - 10$ reported in \citep{Chemerynska2024,Konstantopoulou2024} clearly reflect  the dust accumulation through evolution in previous epochs. Combining this with Eq. (\ref{zmrel}), one can estimate the dust-to-stellar mass fraction of 

\be 
\zeta_\ast={M_d\over M_Z}{M_g\over M_\ast}Z \sim (2-8)\times 10^{-4}{M_g\over M_\ast}\left({M_\ast\over 10^9~\msun}\right)^{0.4}\,.
\ee
Here we have used the fact that solar metallicity $Z_\odot \approx 0.013$. 

Numerical calculations of the long term dust evolution on time scales of $\sim 10$ Gyr by \citet{Asano2013} have demonstrated how dust mass grows over time. With the incorporation of dust destruction by  reverse shocks, it has been shown that within $t\sim 1$ Gyr, the dust-to-gas mass ratio grows as $\zeta_d\sim 10^{-2}(Z/Z_\odot)$. At $t\sim 1$ Gyr, when the metallicity $Z\sim (0.1-0.2)Z_\odot$, this implies a rather low value $\zeta_d\sim (1.3-5.2)\times 10^{-4}$. At later times, $Z/Z_\odot$ asymptotically approaches a level that is an order of magnitude higher than this. The dust-to-stellar mass ratio can be estimated as 
\be 
\zeta_\ast=\zeta_d{M_g\over M_\ast}\sim (1.3-5.2)\times 10^{-4}{M_g\over M_\ast}\,. 
\ee
As is readily seen, the dust-to-stellar mass ratio  still remains below the level observed in the super-dusty galaxies $\zeta_\ast\sim 10^{-2}$, unless the gas mass exceeds the stellar mass by one to two orders of magnitude.

Note that these considerations are only valid  for galaxies in a quasi-steady dynamical state. Galaxies at the epoch of `Cosmic Dawn' are likely to evolve more dynamically than their present-day counterparts. In fact, they are commonly believed to experience bursty energetic events \citep[][]{Dekel2023,Ferrara2024}, that can make their parameter values considerably deviate from those in `average' correlations. 

At the same time, as long as the stellar dust-to-metal yield ratio is small, of order $\le 30$\% \citep[][]{Konstantopoulou2024}, the hypothesis of  dust accumulation from previous epochs would require an unacceptable total stellar mass of previous generations.

\section{Possible solutions}\label{solut} 

It is clear from the discussion above that the `dust budget crisis' would be alleviated if the net SN dust yield $y_{sn}\simgt 1~\msun$. Here we will discuss possible conditions for inhibiting the destruction of SNe injected dust particles by the reverse shock. 

\subsection{Presupernova ionization front, stellar wind and thermal instability} 

As argued above, the dust sputtering behind RS becomes inefficient at lower densities in the ambient gas. As long as the ionization front and the stellar wind from a SN progenitor can evacuate the ambient gas in the HII bubble (Str\"omgren sphere), the sputtering rate decreases. This can reduce the limitations on the total amount of dust survived after processing by ambient plasma particles penetrating into the ejecta, as predicted by  \citet{Nozawa2007,Nath2008,Hirashita2014,Slavin2020,Vasiliev2023}. We will consider this scenario in detail below.

\subsubsection{HII zone.} 

Stellar UV photons ionize and heat surrounding medium. At the initial stages, the ionization front propagates through the ambient medium with a  velocity $v_i\sim \Phi_\ast/4\pi n_0 r^2\gg c_s$, where $\Phi_\ast$ is the number of UV photons per second from the progenitor star, $n_0$ is the ambient density, and $c_s$ is the sound speed. At $t=\tau_r=(\alpha_{b}n_0)^{-1}$ , ionizing photons fill the Str\"omgren sphere, and after a short while, the ionization front transforms into a D-type shock, which sweeps the unperturbed gas and decreases the density in the HII zone. Asymptotically, the density in the HII bubble decreases as $n_{\rm ii}\propto (t/\tau_r)^{-6/7}$ \citep{Spitzer1978,Draine2011}, and can become as low as $n_{ii}\sim 0.03n_0$ at times $t\gg 10^{5}n_0^{-1}$ yr,  \citep[ for recent discussions see in ][ their Figs. 3-4 and 10-12]{Hosokawa2006} and \citep[middle and right panels in Fig. 5 in ][]{Haid2018}. It is worth noting that dust particles with radii $a\geq 0.01~\mu$m survive the radiation field in the conditions that are typical for clustered young stars and planetary nebulae \citep{Nanni2023}.    

\subsubsection{Stellar wind bubble.} 

In addition to UV photons,  stellar wind (SW) also severely evacuates gas from the immediate vicinity of a massive star before the supernova explosion \citep[][]{Gupta2016,Haid2018,Dwarkadas2023,Kourniotis2023}. When the SW begins to act, the gas density in the wind blown bubble falls by $3-4$ orders of magnitude. In other words, when the SN explodes, its ejecta will expand in a medium with $n_w\sim (10^{-3}\hbox{--}10^{-4})n_0$.  An average temperature in the bubble is $T_w\sim 10^8$ K, and one may think that the dust from the following SN can be heavily destroyed when the ejecta gas and the hot bubble percolate through each other. Rough estimates for the sputtering time of particles give $\tau_{sp}\sim 10^8a_{0.1}n_0^{-1}$ yr. Here $a_{0.1}=a/0.1~\mu$m is the dust radius, and the averaged density in the bubble is assumed to be $n_w\sim 10^{-3}n_0$. Therefore, when the effects of gas evacuation by a massive progenitor star are accounted for, the fraction of dust destroyed in the SN ejecta can be considerably reduced unless the mean ambient ISM density exceeds $\langle n_0\rangle \sim 10$ cm$^{-3}$. 

Incidentally, the oppositely directed incoming heat from hot protons, from the wind into the ejecta, proceeds in a diffusive manner and is absorbed within a thin radiatively cooling layer (see details in App. \ref{interf}), with a negligible contribution to dust sputtering. 

\subsubsection{Shielding of dust in dense clumps.}\label{clumps} 

{Observations of local SN remnants demonstrate that the dust nucleated in their ejecta survives the reverse shock even in conditions of a rather dense ambient gas $n_0\sim 1$ cm$^{-3}$. Among the examples are the mentioned above SNRs in the nearby vicinity SN Cas A ($M_d\simeq 0.89~\msun$) and Crab nebula ($M_d\simeq 0.245~\msun$) with the circumstellar density $n_0\sim 2$ cm$^{-3}$ and $\sim 1$ cm$^{-3}$, respectively \citep{Bocchio2016}. } 

{A clumpy structure of the ejecta medium has been suggested as a possible solution of this problem in a series of papers by \citet{Kirchschlager2019,Kirchschlager2023,Kirchschlager2024}. The basic idea is that the reverse shock enters the ejecta with pre-existing dense clumps in it. The clumps are assumed to confine the newly nucleated dust particles. In such conditions the dust particles locked in the clumps are shielded from aggressive influence from  sputtering and grain-grain collision when the reverse shock penetrates the ejecta. As expected, the density contrast --- the ratio of cloud to intercloud densities $\chi=n_c/n_0$ --- is the critical parameter which characterizes the fraction of survived dust. The ejecta with a low and intermediate density contrast $\chi< 600$, shield predominantly large particles with $a\simgt 0.5~\mu$m, whereas at higher gas density in clumps with $\chi\simgt 600$, all grains with $a\simgt 0.01~\mu$m survive. } 

{The presence of clumps in the ejecta is clearly seen in SN 1987A \citep{Dwek2015} in the yet unshocked ejecta. Clumps are present also in the SN Cas A with the reverse shock already entering the central $\sim 3^\prime$ part of ejecta  \citep{Milisavljevic2013,Delooze2017,Milisavljevic2024}. A sample of $\simeq 30$ SNRs in the Milky Way plane revealing ejecta filled with dust clumps within the pulsar wind nebulae are presented by \citet{Chawner2019}. From this point of view, the scenario of dust survival behind the reverse shocks seems promising, although the origin of such clumps does not appear to be obvious. Resolutions that is required to confirm the density contrast needed to protect dust particles from destruction $\chi\simgt 600$ are not currently reachable \citep{Milisavljevic2024}.        
}         

\subsubsection{Shielding of dust by thermal instability.}\label{shield} 

{In addition to the mechanisms that either diminish the ambient gas density or lock the newly formed dust into dense clumps, and thus mitigate the aggressive attack from the RS,} a possible feedback {\it behind} the RS itself can come into play and inhibit its destructive efficiency.  As mentioned above, the reverse shock heats the ejecta up to $T\sim 10^8$ K, and this causes a very efficient thermal sputtering of dust particles  within a characteristic time-scale of $\tau_{sp}\sim 3\times 10^{12}\, a_{0.1}\,n_{nuc}^{-1}$ s, where $a_{0.1}=a/0.1~\mu$m is the grain radius \citep[Eq. 25.14 in ][]{Draine2011}. The ejecta dominated (ED) stage is commonly thought to be non-radiative, in the sense that the duration of the ED phase is shorter than the radiative cooling time. This is the case at the adiabatic stages before and during the nucleation period with $T_{ej}\sim 10^3$ K. 

However, the situation changes when radiative cooling of gas behind the reverse shock is taken into account. In this case, inelastic collisions of hot electrons and ions with dust grains can transfer a great deal of their thermal energy into infrared (IR) emission through the heating of grains. The radiation cooling rate of plasma due to such collisions at $T\sim 10^7-10^9$ K dominates the cooling from inelastic collisions of hot electrons with ions of H and He and metals by more than an order \citep[][ see also \citep{Smith1996,Seok2015}]{Dwek1987}, provided that the dust-to-gas fraction is close to the value in local ISM $0.5\, Z$. 
One can write for the product of the radiative cooling rate and density, as $\Lambda_d \, n\sim 3\times 10^{-21}\bar\zeta_d n$ erg s$^{-1}$ in the temperature range $T=10^6-10^8$ K, where the dust-to-gas mass fraction relative to the mean MW value $\bar\zeta_d=\zeta_d/\zeta_{d,MW}$, assuming $\zeta_{d,MW}=0.01$ \citep{Dwek1987,Seok2015}.     

The gas cooling time is, therefore,  of order (at $T\sim 10^8$ K)
\be \label{cooltime}
\tau_{c}\sim {k_BT\over \Lambda_d n}\sim 4\times 10^{12}\bar\zeta_d^{-1}n^{-1}\,,
\ee 
which gives $\tau_c\simeq 4\times 10^4$ s for the assumed ejecta density of $n\approx 10^8$ cm$^{-3}$. In (\ref{cooltime}) we assumed the  SN dust yield $y_{\rm sn}= 0.1~\msun$, equivalent to $\bar\zeta_d\sim 1$; for  $y_{\rm sn}\sim 1~\msun$, the cooling time $\tau_{c}$ is an order of magnitude shorter. For comparison, the mean dust destruction time within $T\sim (10^6-10^8)$ K is 
\be \label{tspu}
\tau_{sp}\sim (3-5)\times 10^{12}a_{0.1}n^{-1}~{\rm s}\,.
\ee
One therefore can assume that gas in the ejecta behind the reverse shock front cools slightly faster than sputtering of dust particles, and a fraction of dust grains can survive to work as seeds for further possible dust growth in cold and dense regions (clouds). Note that, below $T\sim 10^6$ K, the sputtering time scales as $\tau_{sp}\propto T_6^{-3}$.  

It is worth noting that more recent models of dust growth within the SN ejecta as well as the models of dust destruction by the reverse shocks incorporate possible inhomigeneity (clumpiness) of the ejecta gas \citep[][]{Sarangi2015,Sluder2018}, in contrast to previous calculations based on the approximation of well-mixed gas \footnote{Dust in the Milky Way is believed to be the dominant cooling agent behind shock waves from young SN \citep{Andersen2011}.}.  

Furthermore, the drastic cooling of the gas mixed with dust behind the reverse shocked ejecta leads a thermal instability (TI), as shown below.
This thermal instability of short-wavelength isobaric perturbations can boost the formation of cold and dense clouds behind the reverse shock. Dust particles locked in such clouds would be protected of destructive influence of the hot plasma. The temperature in such perturbations obeys the equation 

\be 
{3\over 2}k_B{dT\over dt}=-\Lambda(T)n\,, 
\ee 
with $n=n_0T_0/T$, where $n_0$ and $T_0$ are the density and temperature at the initial state. The perturbations here are assumed to be isobaric. For a power-law cooling function $\Lambda(T)=\Lambda_0(T/T_0)^\alpha$, the temperature equation can be written as,
\be 
{d\theta\over dt}=-{\theta^{\alpha-1}\over\tau_c}\,,
\ee 
where $\theta=T/T_0$, $\tau_c=3k_BT_0/2\Lambda_0 n_0$. The solution is given by,
\be 
\theta=[1+(\alpha-2)t/\tau_c]^{-1/(\alpha-2)}\,.
\ee 
For $\alpha>2$ this solution describes a slow cooling with temperature at high $t\gg \tau_c$, dropping as $\theta\sim (t/\tau_c)^{-1/(\alpha-2)}$. 
For $\alpha<2$,  cooling becomes faster with the temperature, which drops to zero at a finite time  
\be \label{coolt} 
\theta=[1-(2-\alpha)t/\tau_c]^{1/(2-\alpha)}\,,
\ee
with $\theta=0$ at $t_m=\tau_c/(2-\alpha)$. 

The cooling regime is unstable if a small perturbation of temperature $\delta\theta/\theta$ grows with time. The equation for $\delta\theta$ is 
\be 
{d\delta\theta\over dt}=-(\alpha-1){\theta^{\alpha-2}\over\tau_c}\delta\theta\,, 
\ee
where $\theta=T/T_0$. 
The relative magnitude of  $\delta\theta/\theta$ is governed by the equation 
\be 
{d\over dt}{\delta\theta\over\theta}=(2-\alpha){\theta^{\alpha-2}\over\tau_c}{\delta\theta\over\theta}\,, 
\ee
with the solution 
\be \label{dtrel}
{\delta\theta\over\theta}= \left({\delta\theta\over\theta}\right)_0[1-(2-\alpha)t/\tau_c]^{-1}\,, 
\ee 
which grows hyperbolically at $t_{m}=\tau_c/(2-\alpha)$, at which point the temperature $\theta$ vanishes. This tentative analysis encourages further numerical consideration in order to confirm thermal instability as a mechanism that is capable of shielding growing dust from efficient sputtering.    
     
\subsection{Numerical model of thermally unstable gas behind the reverse shock}\label{tinu} 

In order to illustrate how the thermal instability driven by dust radiation cooling develops behind the reverse shock (RS), we consider a simplified model, representing the growth of small perturbation in a hot medium in the ejecta region after the RS passes through it. We consider first the case of  a spherically symmetric perturbation with the help of a 3D numerical procedure, as  briefly described below.  We begin with a discussion of the cooling rate in the presence of dust.

\subsubsection{`Dust cooling'}\label{dstcool} 

Let us assume the ejecta mass to be $M_{ej}=10~\msun$, contained within a spherical region of radius $R_{ej,0}=10^{3}R_\odot$ cm. The corresponding initial gas density $n_{ej}\simeq 7\times 10^{15}M_{10}R_{13}$ cm$^{-3}$ as in Eq (\ref{nej}), and the initial gas temperature to be $(0.1-1)\times 10^8E_{51}M_{10}^{-1}$ K, Eq (\ref{teja}). At the time of nucleation of solid particles, $\sim 400\hbox{--}600$ day, the gas temperature in ejecta is $T\simeq 3000$ K, and the corresponding gas density $n_{nuc}\simeq 10^{8}$ cm$^{-3}$ as in Eq. (\ref{nnuc}). The nucleation proceeds rapidly and within a couple of months the ejecta contains  $\sim 0.1-3\msun$ of dust  \citep{Todini2001,Bianchi2007,Cherchneff2009,Cherchneff2010,Sarangi2013,Sarangi2015,Marassi2015,Marassi2019}. We, therefore, set approximately $n=10^8~{\rm cm}^{-3}\sim n_{nuc}$. 

\begin{figure} 
\center
\includegraphics[width=8cm]{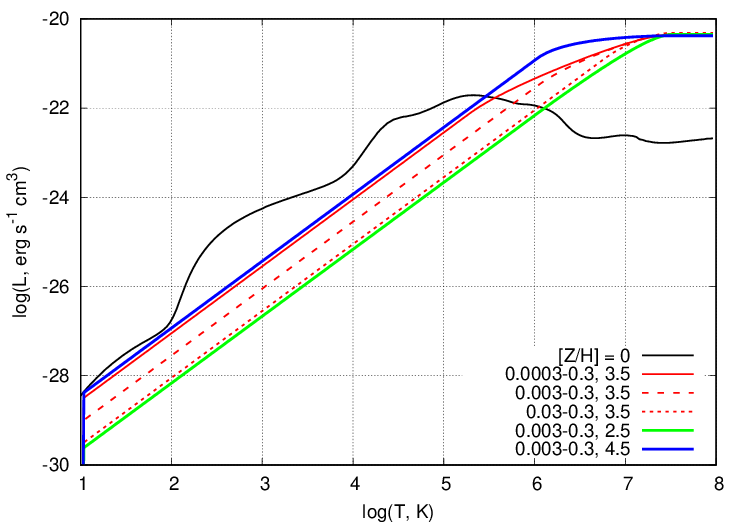}
\caption{The gas cooling function adopted in simulations: black solid line show the gas cooling function without contribution from dust; dust cooling function is shown by color lines for dust-size distributions within $a=[a_1:a_2]$ -- the first column, and the power-law index $p$ -- the second column in the legend. The dust-to-gas mass ratio is $\zeta_d=0.01$ for all curves.   
}
\label{coolf}
\end{figure}

The gas cools radiatively due to a collisional transfer of energy from hot electrons to dust particles, which subsequently re-emits in the infrared. The rate of dust heating under collisions from hot thermal electrons is \citep[][]{Dwek1987}  

\be \label{dh}
{\cal H}(T,n)=\left({32\pi\over m_e}\right)^{1/2}n_e(kT)^{3/2}C\int\limits_{a_1}^{a_2}a^{2-p}h(a,T)da\,,
\ee
where $m_e$ is the electron mass, $a$ is the dust grain radius, $h(a,T)$, the dust heating efficiency, approximated from \citep[][see Appendix \ref{hat} below]{Dwek1987}, and the dust size distribution function (DSD) is assumed to be in the form $n(a)=Ca^{-p}$ in the range $a=[a_1:a_2]$ such that the dust-to-gas mass ratio $\zeta_d$ is kept fixed. With this assumption the normalizing coefficient is defined as 

\begin{figure*} 
\center
\includegraphics[width=16cm]{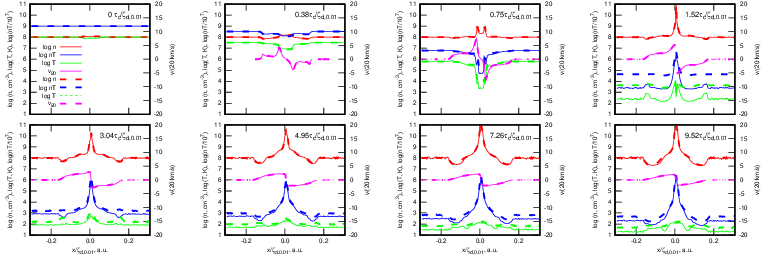}
\caption{Thermal evolution of an isolated spherical perturbation (`clump'). In the panels from top to bottom, each panel represents 1D profiles of density (red), temperature (green), pressure (blue) and radial velocity (pink) along a diagonal segment of the computational zone, for different times as indicated on top. The $x$-axes refer to the vertical projection of the segment. Thin solid lines depict the model with $\zeta_d=0.01$ ($\bar\zeta_d=1$), while thick dashed are for $\zeta_d=0.03$ ($\bar\zeta_d=3$). The initial size (diameter) of the perturbation is approximately equal to the cooling length $\lambda\sim \lambda_c=2\times 10^{12}\bar\zeta_d~{\rm cm}$; time in legends is shown in units of day/$\bar\zeta_d$,  $x$-axis scale with AU/$\bar\zeta_d$, and the dust-to-gas ratio is in the units of the local ISM. It is seen that at a time-scale of 3 days ($\simeq  12\tau_c$), density and temperature perturbations grow nearly isobarically with slowly varying pressure; $\tau_c\simeq 0.25$ d for $T=10^8$ K, $n=10^8$ cm$^{-3}$. The `conservative' case for cooling function $\Lambda(T)$, with MRN dust shown in Fig. \ref{coolf} in red, is adopted in this model.  
}
\label{evolsph}
\end{figure*}

\be \label{coolnorm}
C={3(4-p) \rho \zeta_d\over 4\pi\bar\rho(a_2^{4-p}-a_1^{4-p})}\,, 
\ee 
with $\rho=\mu m_{\rm H}$ being the gas density, $\bar\rho=2.2$ g cm$^{-3}$, the density of the  dust grain material. Energy gained by dust particles from thermal plasma electrons in Eq. (\ref{dh}) is equivalent to the cooling rate of plasma: $\Lambda(T)n\equiv {\cal H}(T,n)$. This process is referred to as `dust cooling' \citep[][]{Dwek1987}.    

The total cooling function in the temperature range of ambient gas $T=10-10^8$ K is shown in Figure \ref{coolf} for different models of dust-size distributions as described in the figure legend. It is clearly seen that dust cooling dominates at $T>3\times 10^5$ K. At the high temperature end $T\sim 3\times 10^6-10^8$ K, the dust cooling function can be approximated by a power-law with a slope $\alpha\approx 0-0.5$. As shown above, in this case the perturbation develops with the overdensity growing as $\delta n/n\propto |\delta T/T| \to \infty$. 

\begin{figure} 
\center
\includegraphics[width=8cm]{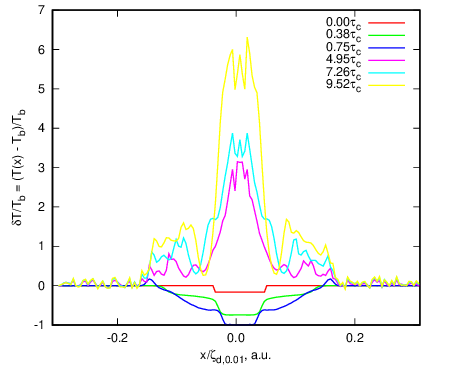}
\caption{The plot shows the evolution of the relative temperature perturbation $\delta T(t)/T(t)$. Despite a general similarity between the peak value and the analytic solution in Eq. (\ref{dtrel}), there is an obvious difference, which is connected with the adiabatic heating and cooling by the acoustic wave propagating from the boundary and back in the process of a relaxation of the perturbation.   
}
\label{evolnoniso}
\end{figure}

\subsubsection{Isolated spherical perturbation}\label{isolatesph} 

In our model we start with a localised isobaric perturbation $\delta n/n=-\delta T/T$ in a uniform gas. The dynamics of the instability is determined by the interrelation between characteristic time and spatial scales that govern the process. Besides the dynamical time, $t_d$  
and length $\lambda_d=c_s t$, the cooling time and length scale as follows:

\be \label{tcool}
\tau_c={3\over 2}{k_BT\over \Lambda n}\sim 2\times 10^4~{\rm s},~\lambda_c=c_s\tau_c\sim 2\times 10^{12}~{\rm cm}\,.
\ee 
Here we assumed $n\sim 10^8$ cm$^{-3}$, $T\sim 10^8$ K, $\Lambda=\Lambda(T=10^8~{\rm K})\sim 7\times 10^{-21}$ erg cm $^3$ s$^{-1}$. Thermal instability grows when the perturbation size exceeds the Field length \citep{Field1965} 

\be \label{field}
\lambda_{F}=\left({k_B\kappa T\over \Lambda(T)n}\right)^{1/2}\sim 3\times 10^9~{\rm cm}\,,
\ee
which is the thickness of a layer $\Delta\ell$ where the heat flux between the hot ambient gas and the colder gas inside the perturbation is exactly balanced by radiative cooling. Within the interface, one has the balance between $k_B\kappa \nabla nT\sim \Lambda(T)n^2\Delta\ell $. The numerical value in the r.h.s. of Eq. (\ref{field}) is obtained for the  value of physical electron conductivity of $\kappa_e\approx 6.7\times 10^{-6}T^{5/2}$ cm$^2$ s$^{-1}$ \citep{Ferrara1993}, assuming the values for temperature and density mentioned earlier. However, in our model, numerical diffusivity $\kappa_n\sim c_s \Delta x/3\sim 3\times 10^{18}$ cm$^2$ s$^{-1}$ is much higher than $\kappa_e$, such that the equivalent Field length is dominated by numerical effects 

\be 
\lambda_F^n\sim 10^{11}~{\rm cm}\,. 
\ee 
In a thermally unstable medium, only perturbations with the wavelength $\lambda>\lambda_F$ grow. 

In order to avoid strong hydrodynamical motions that can interfere with a smooth development of the instability, in the model shown in Fig. \ref{evolsph}, the perturbation size is set to be $r_p=\lambda_c=2\times 10^{12}$ cm, for which one has $r_p\gg \lambda_F^n$. In these conditions, the Field length determines the thickness of the interface layer between the growing cool cloudlet and the hotter ambient gas. 

{At $t=0$, a top-hat perturbation of density and an inverse-hot-hat perturbation of temperature with magnitudes $\delta n_0/n_0=0.2$ and $\delta T_0/T_0=-0.2$ is applied at the center of the computational domain, where one has $n_0=10^8$ cm$^{-3}$, $T_0=10^8$ K. The size of the perturbation is taken to be larger than the Field length, $\lambda\sim \lambda_c\gg \lambda_F^n$, as mentioned above. In this case the heat flux from the ambient hot gas is absorbed into a thin layer at the perturbation boundary, and isolates the perturbation from `evaporation' due to a destructive diffusive heating\footnote{In our numerical model the heat flux is due to numerical thermal diffusivity.}.} 

\begin{figure*} 
\center
\includegraphics[width=17cm]{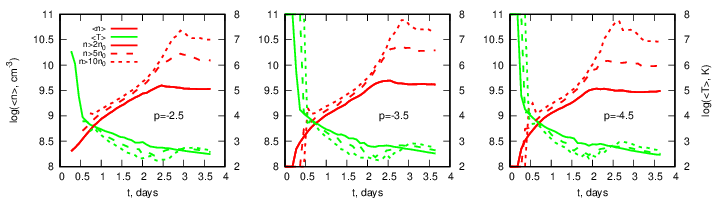}
\caption{Examples of growing thermal isobaric perturbations in a hot plasma behind the reverse shock wave under the dust cooling. The curves represent the overdensity $\delta n/n$ and the corresponding $\delta T/T$; 1 day$\simeq 8.6\times 10^4~{\rm s}\simeq  4\tau_c$, with $\tau_c$ being the cooling time in the initial state: $T=10^8$ K, $n=10^8$ cm$^{-3}$. 
 The initial size (diameter) of the perturbation is $2r=\lambda_c$. A power-law dust size distribution $n(a)\propto a^{-p}$ with $p=2.5$ (left), $p=3.5$ (middle) , $p=4.5$ (right panel), and $a_{1}=0.003~\mu$ and $a_2=0.3~\mu$m, the dust-to-gas ratio in all models is fixed at $\zeta_d=0.01$, and the cooling rate is assumed to be as shown in Fig. \ref{coolf}. Green lines show the temperature, whereas red lines refer to density. 
 }  
\label{tins}
\end{figure*}

{Gas in the perturbed region rapidly cools  within $t\simlt \tau_c$, closely following the hyperbolic solution (\ref{dtrel}). At $t=0.38\tau_c$, the temperature of ambient gas falls approximately by $|\Delta T|_{amb}\simeq  0.65T_0=6.5\times 10^7$ K. Within the perturbed region, temperature falls by $|\Delta T|_{pert}\sim 0.3T_0=3\times 10^7$ K with respect to the ambient temperature. This leads to a trough in the temperature profile with $T_{pert}\sim 10^7$ K and $T_{amb}\simeq 3.5\times 10^7$, as seen in the second upper panel in Fig. \ref{evolsph}.}

{At later times, the difference between the numerical and the simplified isobaric solutions (Eq. (\ref{dtrel})) becomes evident. While the temperature in the perturbed region drops considerably within  $t=0.38\tau_c/\bar\zeta_d$, the ensuing acoustic wave can cover only a small distance inside the perturbation $\delta r\sim 4\times 10^{11}~{\rm cm}\ll \lambda_c$, (the second upper panel in Fig. \ref{evolsph}). This means that  acoustic waves are not capable of redistributing the density in the perturbed region isobarically, i.e., $\delta n/n\simeq -\delta T/T$. Hence the relative increase in the density is only $\delta n/n\simeq 1.5\delta n_0/n_0\simeq 0.3$, considerably smaller than the decrease in the temperature. As a consequence, the pressure inside the perturbed region approximately follows the decrease in temperature, and forms an over-pressured region at the boundary, $P_{ex}\gg P_i$ (where the subscript $ex$ refers to external, $i$ refers to internal). This difference in pressure $\Delta P=P_{ex}-P_i$ drives the sound wave inwards with a speed $|\delta v|\sim 60-80$ km s$^{-1}$, which is $\simeq 8$\% of the sound speed in ambient gas (second upper panel in Fig. \ref{evolsph}). }

{This slow relaxation of pressure in the perturbed region explains the numerically obtained deviation of the density and temperature  evolution from the isobaric mode assumed above in Sec. \ref{shield}. The inward flow from the perturbation boundary with the velocity $v_{in}\sim 60-80$ km s$^{-1}$, reaches the center within $t\simeq 1.5\times 10^5$ s, which is an order of magnitude larger than $\tau_c$. Therefore, it takes a longer time for the inflow to considerably increase the density in the innermost parts of the cooling perturbed gas. As a result, {the internal pressure} {$P_i$ remains unrelaxed and the difference $\Delta P$ increases following}  further cooling inside the perturbed volume. The external pressure $P_{ex}$ changes at a slower rate than this and remains considerably higher, $P_{ex}\sim 100 P_i$ by $t=0.75\tau_c/\bar\zeta_d$. As the temperature decreases ($T\simlt 10^5$ K outside and $T\simlt 3\times 10^3$ K inside the perturbed region), the inflow becomes supersonic and generates an implosive compression of outer layers. This leads to an increase in density  at the boundary by factor of $\sim 4$ higher than in the central part of the cooled region. The temperature in the central part drops to $T\sim 3\times 10^3$ K as seen in the third upper panel in Fig. \ref{evolsph}.} 

Conditions in the gas confined within perturbations of shorter wavelength, $\lambda\simlt\lambda_c$, turn out to be more favorable for survival of dust particles. In this case the perturbations do not show strong implosive motions, and during the initial $t\sim \tau_c$ they develop nearly isobarically with $nT\approx n_0T_0$. As a result, the cooling time is half of its value in the isochoric process. 

{Further development qualitatively changes the overall dynamics. We show the temperature profiles at different times in Figure \ref{evolnoniso}.  At $t\simgt \tau_c$, radiation losses decrease by two orders of magnitude. At temperatures below $<10^5$ K, cooling becomes insufficient to counteract the adiabatic heating due to a strong converging flow. This transition from a fast radiation cooling to adiabatic compression can be observed in Fig. \ref{evolnoniso}. At $t\simgt \tau_c$ (c.f., Eq. \ref{dtrel}), the compressive adiabatic heating driven by the implosive shock wave comes into play and dominates at later stages. The sound speed in the perturbed region at longer times ($t\simgt 5\tau_c$) falls below $c_s\sim 1$ km s$^{-1}$. It is reasonable to think that during subsequent evolution, at time scale $t\simgt \lambda_c/c_s\sim 0.5$ yr, this cold compressed region will decay to diverging sound waves.   
}

Figure \ref{tins} depicts the evolution of the average properties of the gas across the entire cooling perturbation. In the process of development of the thermal instability within an isolated clump, the gas rapidly transfers to a cold and dense state within the cooling time: the average temperature drops (Fig.  \ref{tins})  below $\langle T\rangle \simlt 10^5$ K, while the average density increases at a slower rate. It grows by an order of magnitude in time scales of $t\sim 4\tau_c$, as shown in Fig. \ref{tins} by the red solid line. The denser regions are generally colder, in agreement with the distributions shown in spatial profiles in Fig. \ref{evolsph}. 

The three panels represent the average evolution of perturbed region with increasing slopes of the dust size distribution: left to right from $p=2.5$, to $p=3.5$ and $p=4.5$ with equal amount of the total dust mass. As the curves show, a steeper slope of $p$ corresponds to a faster initial cooling and a sharper compression.  
As is seen in Fig. \ref{coolf}, on average, steeper slopes of dust size distributions $p$ result  in an increase of the cooling function $\Lambda(T)$ in the range $T=10^6-10^8$ K, and make its dependence on $T$ flatter. This can be understood from the fact that in the case of a steeper slope of $p$, the  size distribution has an enhanced contribution of small particles to `dust cooling', causing the thermal instability to grow faster.

\subsubsection{Conditions for dust survival} 

In order to estimate the fraction of dust mass that survives in cold dense clumps under thermal instability, we assume the dust to be strongly coupled to the gas component. Such an assumption is justified by the fact that the momentum exchange between dust and gas particles is efficient. It is seen from the comparison of the dust `stopping' (friction) time \citep{Baines1965,Draine1979a} 
\be \label{stotime}
\tau_{st}\sim {\rho_m a\over \rho_g \sigma_t}\,,
\ee
with the dynamical (crossing) time $\tau_{cr}=r_p/c_s$ and the cooling time $\tau_{c}$ (Eq. \ref{tcool}. Here $\sigma_t\sim 10^8T_8^{1/2}$ cm s$^{-1}$ is the thermal velocity. It turns out that the stopping time $\tau_{st}\sim 10^3 a_{0.1} n_8 T_8^{-1/2}$ s (where $n=10^8 n_8$ cm$^{-3}$)  is much shorter than the crossing and the cooling time scales: $\tau_{cr}\sim 10^5 r_{\rm au}T_6^{-1/2}$ s and $\tau_c\sim 2\times 10^{4}n_8^{-1}$ s, correspondingly. {This means that dust particles remain tightly coupled to the ambient gas\footnote{{The distance that dust particles cross during one cooling time ($\tau_{c}$) under random (Brownian) motions can be estimated as $\langle\Delta x^2\rangle^{1/2}\sim [2k_BT\tau_c/\pi a^2\rho \sigma_t]^{1/2}\sim 10^7T_8^{1/4}a_{0.1}^{-1}n_8^{1/2}$ cm, which is  7 orders of magnitude smaller than the computational cell size.}}.} 

\begin{figure} 
\center
\includegraphics[width=8cm]{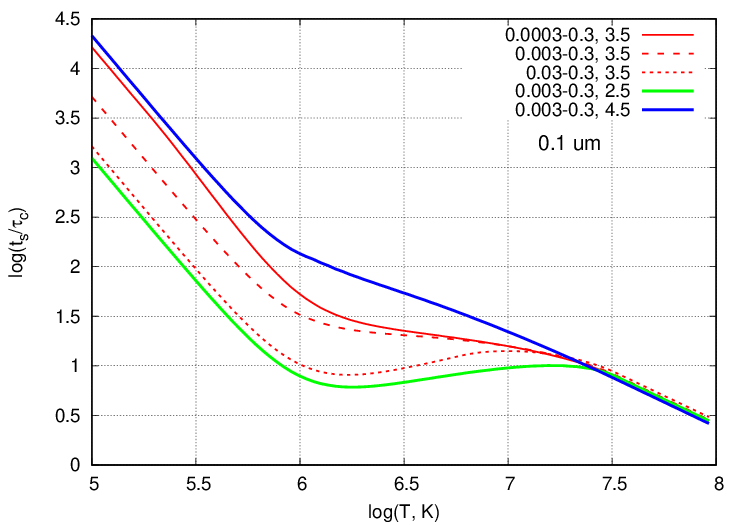}
\caption{The ratio of sputtering to cooling time scales $\tau_{sp}(a)/\tau_{c}$ as a function of the ambient gas temperature for $a=0.1~\mu$m. The ratio is shown for the cooling operated by the cooling functions shown in Fig. \ref{coolf}; colors and line types are as in Fig. \ref{coolf}.   
}
\label{sputoco}
\end{figure}

In such conditions, the local ratio of the cooling to the sputtering times $\tau_c/\tau_{sp}$ can be an approximate indicator of dust survival: the exponent $e^{-\tau_c/\tau_{sp}(a)}$ will roughly characterize the fraction of destroyed dust of a given radius $a$ in the cooling time scale $\tau_c$. Since $\tau_c\propto T$ (Eq. \ref{tcool}) and $\tau_{sp}$ is nearly invariant at $T\sim 10^6-10^8$ K (Eq. \ref{tspu}), the decrease in the ambient temperature by factor of $\sim 2$ considerably weakens the destructive effect of sputtering.  

{The sizes of dust grains evolve synchronously with the ambient gas temperature in the course of formation of an isolated cloudlet. The dust grain size $a(T)$ and gas temperature evolve according to  the following equations: }

\be \label{avt}
{da\over dt}=~-{a\over \tau_{sp}} \,, \qquad {dT\over dt}=~-{2\over 3k_B}\Lambda(T)n \,,
\ee
and 

\be \label{tvert}
{dT\over dt}=~-{2\over 3k_B}\Lambda(T)n \,.
\ee
This can be reduced to,
\be 
{da\over dT}={\tau_c\over \tau_{sp}}{a\over T}\,,
\ee  
where the sputtering time 

\be\label{sput}
\tau_{sp}(a,T)\approx 3\times 10^{17}e^{0.7/T_6}(1+T_6^{-3})\, a \, n^{-1}~ {\rm s}\,,
\ee 
{and where $10^6T_6=T$ K.  
The exponential factor $e^{-0.7/T_6}$ in the sputtering rate is added in order to extrapolate the power low approximation $(1+T_6^{-3})^{-1}$ which is valid at $T_6>1$ to lower temperatures \citep[Fig. 25.4 and Eq 25.13 in ][]{Draine2011}, see also in \citep{Draine1979a}. The cooling time }

\be 
\tau_c(T)\sim {3\over 2}{k_BT\over \Lambda(T)n}\approx 3\times 10^{12}\bar\Lambda(T)^{-1}{T\over 10^8}n^{-1}~{\rm s}, 
\ee
{where $\bar\Lambda(T)=\Lambda(T)/ \Lambda(10^8)$. 
We use a dust cooling function in the temperature range of $T=10^6-10^8$ K, as  shown in Fig. \ref{coolf} for MRN dust size spectrum with $a=0.0003-0.3~\mu$m assuming the MW dust fraction to be $\bar\zeta_d=1$. }

At high temperatures of the ambient gas $T\simgt 3\times 10^6$ K, the dust radius decreases approximately as 
{
\be \label{athigh}
{a\over a_0}\sim 1-{\tau_c\over \tau_{sp}(a)}\left[1-\left({T\over T_0}\right)\right]\,. 
\ee }
This equation  clearly shows that the sputtering rate increases (or decreases) in a cooling medium depending on whether the times ratio $\tau_c/\tau_{sp}(a)$ is larger (or smaller) than unity. In the first case $\tau_c/\tau_{sp}(a)>1$, the hot environment stays longer than that is needed for sputtering to operate. On the contrary, in the later case $\tau_c/\tau_{sp}(a)<1$, the ambient gas cools before the sputtering comes into play.

The difference between the sputtering and the cooling times is illustrated in Fig. \ref{sputoco} in terms of  the ratio $\tau_{sp}(a)/\tau_{c}$ versus gas temperature for a dust particle with radius $a=0.1~\mu$m. As it is readily seen, the sputtering rate of dust of this size is always slower than the cooling rate of the gas. Since $\tau_{sp}(a)\propto a$, the ratio $\tau_{sp}(a)/\tau_{c}$ for smaller particles is proportionally lower and can become less than unity at higher temperatures. It is also important to stress that in the high temperature range, $T>10^6$ K, the ratio $\tau_{sp}(a)/\tau_{c}$ falls nearly as $\propto T^{-0.75}$ due to an enhanced cooling, while it increases (because $\tau_{sp}\propto e^{0.7/T_6}$) towards lower temperatures, $T<10^6$ K, due to an inhibited sputtering.

\subsubsection{A Kolmogorov perturbation field} \label{kolmog}

Let us consider a more realistic model with perturbations spread over the entire computational zone. For this purpose we model the perturbation field with a Kolmogorov  spectrum $E_k\propto k^{-5/3}$. The amplitude of the density perturbations was assumed to satisfy a lognormal distribution with  dispersion $\sigma=0.2$ (equivalent to $|\delta n/n|\simeq 0.22$ in the spherical model); the perturbations were assumed to be isobaric: $\delta T/T=-\delta n/n$. In this regard, we follow the prescription suggested by \citet{Lewis2002}.  

\begin{figure*} 
\center
\includegraphics[width=16cm]{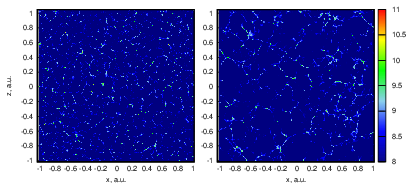}
\caption{The two slices of density and temperature distributions across the computational domain over the $z=0$ plane represent the models with widespread distributions of the perturbation. On the {\it left} the maximum wavelength of the perturbation spectrum is approximately the cooling length $\lambda_{max}=\lambda_c$, while on the {\it right}, it is twice of the cooling length $\lambda_{max}=2\lambda_c$. The snapshots are shown at a time of $t=0.18$ day. The amplitudes of pressure and velocity generally decrease with time, indicating a tendency of the perturbed gas to relax to a new equilibrium. However, as can be observed in Fig. \ref{evoltins}, the system relaxes slower than in the case of an isolated spherical clump, because of the wider range of wavelengths and their mutual interactions and excitations of new wave modes.  
}
\label{2image}
\end{figure*}

\begin{figure*} 
\center
\includegraphics[width=16cm]{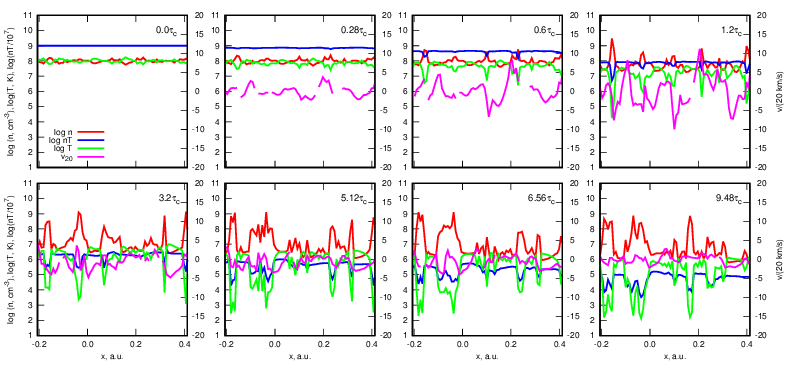}
\caption{{Same as in Fig. \ref{evolsph} for perturbations spread over the computational zone of 2$\times$2 AU. The panels from left to right and from upper to lower row represent 1D profiles of density (red), temperature (green), pressure (blue) and radial velocity (pink) along the diagonal of a square segment of the computational zone, as indicated on left panel of Fig. \ref{2image}, for different times as indicated on top; $x$-axes shows the vertical {projection} of the diagonal. The maximum wavelength of the perturbation spectrum is approximately equal  to  the cooling length $\lambda=2\times 10^{12}~{\rm cm}\sim \lambda_c$.}
}
\label{evoltins}
\end{figure*}

The dynamics of clumpy perturbations in this model differs from that one finds in an isolated sphere, because of mutual distortions from partially overlapping growing clumps in the vicinity. Nevertheless, the general features remain similar to those demonstrated by the spherical perturbation in Fig. \ref{evolsph}: regions with a growing density are strongly connected with those with a decreasing temperature, as can be observed in comparing the left and right panels on Fig. \ref{2image}. In the left panels these regions are presented by dense and cold walls of a web-like structure. A similar structure is seen also in the right panels, but with a stronger contrast of density and temperature distributions because of a more smooth development of the thermal instability in cases with $\lambda_m\sim \lambda_c$ in the left panel. The dynamical differences between these two regimes follow the discussion above (Sec. \ref{isolatesph}). The 1D profiles of physical variables along a diagonal of a selected segment of the entire 2D zone presented in Fig. \ref{2image} are shown in Fig. \ref{evoltins}. 

{As mentioned above the maximum wavelength of the perturbation in the field of left panels in Fig. \ref{2image} and in 1D profiles in Fig. \ref{evoltins}, is close to $\lambda_c\simeq 2\times 10^{12}$ cm ($\approx 0.3$ AU), and the magnitude $\sigma\sim 0.2$ is comparable to the model of the isolated spherical perturbation. It is seen that in a third of the cooling time $t=0.3\tau_c\simeq 0.07$ d, the perturbations become close to isobaric and   show a considerable growth in the magnitude of $\Delta n/n\sim |\Delta T/T| \sim 1.0$. At this time, a small drop of pressure $|\Delta P/P|\sim 0.07P_0\sim 0.7$ dyne cm$^{-2}$ at the interfaces between the `clouds' with $\delta\rho>0$ and the `inter-cloud gas' with $\delta\rho<0$ provokes random gas motions. In contrast to the isolated spherical model, velocity perturbations in the random field not only support compression of cooling gas in the clumps, but also partially work towards adiabatically heating the  low density `inter-cloud' gas. The magnitude of diverging and converging velocity perturbations in the field presented in Fig. \ref{evoltins} is of order $|\delta v|\sim 50$ km s$^{-1}$. The kinetic energy density in the perturbation is $\epsilon_k\sim \langle\rho\rangle\delta v^2/2$. The corresponding energy flux into a given region of radius $r$ can be estimated as }

\be 
{\cal F}_k\sim 4\pi r^2\epsilon_k |\delta v| \,. 
\ee
{When this energy flux is applied to the inter-cloud gas of radius $r$, it can support radiative energy loss in the volume provided that} 

\be 
{\cal F}_k={4\pi\over 3}r^3\Lambda(\langle T\rangle)\langle n^2\rangle \,, 
\ee 
resulting in 

\be 
r\sim 3{\epsilon_k|\delta v|\over \Lambda(\langle T\rangle)\langle n^2\rangle}\sim 0.1~{\rm AU}\,, 
\ee
{with the r.h.s. being estimated at $\langle T\rangle\sim 3\times 10^3$ K, $\langle \rho_i\rangle\sim 3\times 10^6$ cm s$^{-3}$, and $|\delta v|\sim 50$ km s$^{-1}$, typical for the `inter-cloud' gas. This demonstrates that the inter-cloud gas can be supported in a quasi-static equilibrium with the average parameters as in Fig. \ref{evoltins} with characteristic sizes of $\sim 0.1-0.2$ AU. Similar estimates for the `cloud' gas with $\langle T\rangle \simlt 10^3$ K would require further compression, with densities of order $\delta n\simgt 10^{10}$ cm$^{-3}$.} 

{This determines the main difference between the dynamics in models with the single spherical perturbation and the one with randomly spread perturbations. In the first case, the pressure drop at outer layers of the perturbed region generates a supersonic velocity localised around the center. This results in a strong implosion with a rapid increase of central density accompanied with a fast overcooling below $T<1000$ K. In the second case, the velocity field driven by randomly distributed pressure gradients acts as the energy source that stimulates transition of the hot inter-cloud gas into the cold and dense clouds. This process supports a quasi-equilibrium state in which the mass of cold gas slowly increases due to a gradual accretion from the unstable hot gas.    } 
 
Under these circumstances, the dust destruction regime  differs from the one operating in an idealized spherical perturbation. This difference is illustrated Fig. in \ref{evoltins} by a relatively hotter region  with $T\simgt 10^6$ K and the size $\Delta x\sim 0.3$ AU located at around $x\simeq 0.0$ AU. The  diverging velocity around this region with amplitude $v\sim 30$ km s$^{-1}$ will disintegrate it in  $t\sim 0.5$ yr. This is much longer that the time $t\sim 6\tau_c\sim 10^5$ s between the left and right panels in Fig. in \ref{evoltins}, and the entire evolutionary time of the spherical perturbation in Sec. \ref{isolatesph}. Effectively, the supersonic velocity motions generated under thermal instability in a radiatively cooling medium with random perturbations support the perturbed gas at a higher mean temperature and its peak values. As a consequence, the dust particles become locked in the destructive environment for a longer duration, and the net destruction is  larger as compared to conditions in a freely cooling isolated perturbation.

\subsubsection{Dust survival fraction}\label{2flu} 

{The analysis based on the comparison of the cooling and sputtering time scales described above can qualitatively illustrate  only the generic interrelations between the two processes behind the reverse shock. More precise estimates of dust destruction require a full 3D numerical consideration, with dust particles as a separate fluid connected to the gas collisionally, as described in \citep{Vasiliev2023}. For this purpose we use the total variation diminishing (TVD) approach to ensure high-resolution capturing  of shocks and inhibition of unphysical oscillations, with the Monotinic Upstream-Centered Scheme for Conservation Laws (MUSCLE)-Hancock scheme, and  Haarten-Lax-van-Leer Contact (HLLC) method as approximate Riemann 
solver \citep{Vasiliev2015,Vasiliev2017,Vasiliev2023}.     
}

\begin{figure} 
\center
\includegraphics[width=7.5cm,angle=00]{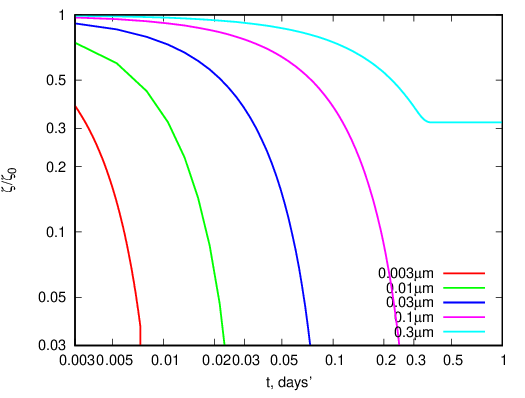}
\includegraphics[width=7.5cm,angle=00]{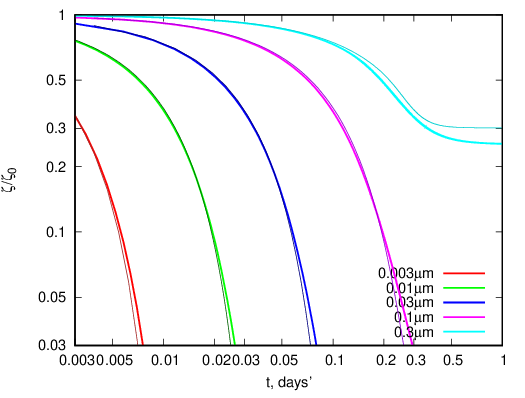}
\caption{Evolution of the dust-to-gas mass ratio for single size dust particles. {\it Left panel:} illustrates the case of the isolated spherical perturbation shown in Fig. \ref{evolsph}. {\it Right panel:} is for the randomly perturbed field in a hot environment (Fig. \ref{evoltins}). The results are for a set of grain radii $a=0.003,~0.01,~0.03,~0.1,~0.3~\mu$m as shown in the legend, the initial temperature $T_0=10^8$ K. Thin lines in {\it lower panel} represent $\sigma=0.1$, thick lines are for $\sigma=0.2$.  
}
\label{2fluspu}
\end{figure}

{For the initial conditions, we used the density and the temperature as expected in the ejecta in the nucleation region behind the RS as in (Eq. \ref{nnuc}) 
\be 
n_{nuc}\simeq 10^{8} M_{10}^{5/2}R_{0,3}^{-3}E_{51}^{-3/2}~{\rm cm^{-3}}\,,
\ee 
and in (\ref{tnuc}) 
\be 
T_{rs}\sim 2\times 10^8{E_{51}\over M_{10}}~{\rm K}\,.  
\ee  
In these conditions the stopping time (Eq. \ref{stotime}) 
\be 
\tau_{st}\sim {\rho_m a\over \rho_g \sigma_t}\sim 10^3a_{0.1}n_8T_8^{-1/2}~{\rm s}\,, 
\ee }
which is much shorter than the cooling and sputtering time scales, such that the dust particles are strongly coupled to gas, and thus can be treated as a fluid being coherently co-moving in given thermal conditions connected with the Eulerian gas, see \ref{explic}. The results are show in Fig. \ref{2fluspu}, where the dust-to-gas mass ratio for each particle size $a$ $\zeta_d(a)$ relative to their initial values $\zeta_{d,0}(a)$, are presented for {the two considered models:  isolated spherical perturbation (left panel), and  Kolmogorov random field with $\sigma=0.1$ and $0.2$, $\lambda_{\rm max}=\lambda_c$, as described in Sec \ref{kolmog}, and Fig. \ref{2image}, with the initial temperatures $T_0=10^8$ K (middle panel), and $T_0=3\times 10^7$ K (dashed lines in the right panel).} 

{It is readily seen from a comparison of the upper and lower panels for particles of $a=0.3~\mu$m, that the spherical perturbation is more sparing to dust: for perturbations of equal initial amplitude $\delta T/T\simeq 0.2$, the survived dust fraction in the first case is $\sim 20$\% larger than in the second case with random perturbation field. Even for a lower amplitude of random perturbations $\delta T/T\simeq 0.1$ this fraction remains by $\sim 10$\% higher in the first case than in the second one. This difference is connected with the fact that randomly perturbed gas is kept at higher temperature $\langle T\rangle \simeq 10^3$ K with $T\sim 10^4$ K peaks in a considerable fraction of the volume, see Fig. \ref{evoltins}.        
}

\subsubsection{{Self-consistent cooling function}}\label{scfcool}

As seen from Fig. \ref{2fluspu}, small dust particles $a\simlt 0.3~\mu$m are destroyed within $t\sim \tau_c$ ($\simeq 0.25$ d), when the gas temperature behind the RS, $T_{rs}\sim 3\times 10^7$ K, still remains efficient in sputtering. The survival mass fraction of all particles with radius $a\leq 0.1~\mu$m falls below $\sim 0.03$. This results in a proportional decrease of the dust cooling function nearly proportional to $\zeta_d(a\leq 0.1~\mu$m), as long as the dominant contributions to dust cooling comes from the small size particles. Consequently, the ratio  $\tau_c/\tau_{sp}$ increases with accompanied progressive enhancement of dust destruction. 

Another consequence is a weakening of the thermal instability due to a steepening of the temperature dependence of the cooling function. As seen in Fig. \ref{coolf}, a flattening of the dust size distribution (green line) leads to a steepening of the cooling function with temperature $\Lambda\propto T^\alpha$ with $\alpha\to 2$. As a result, the characteristic time of thermal instability grows to $t_{TI}\sim \tau_{c}/(2-\alpha)$ and becomes longer than the cooling time of the ambient gas. 

\begin{figure} 
\center
\includegraphics[width=7.5cm,angle=00]{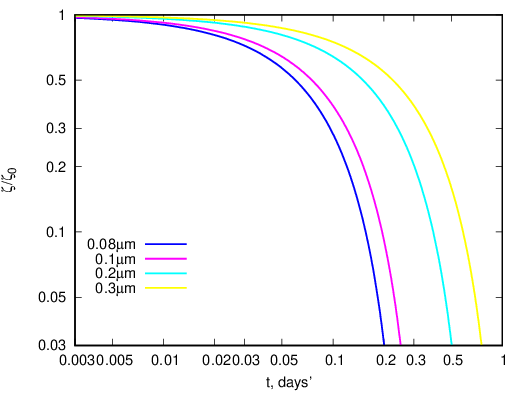}
\caption{Evolution of the dust-to-gas mass ratio for single-sized dust particles with the cooling function adapted at every time-step to dust destruction. As in the previous model, with a perturbed field shown in Fig. \ref{evoltins}, the initial temperature is assumed $T=10^8$ K,  $\sigma=0.1$. 
}
\label{2fluss}
\end{figure}

In order to account for this effect in our calculations of dust destruction, we corrected the cooling function at each time-step by the decreasing dust mass fraction $\zeta^{i}_d(t)$. The result shown in Fig. \ref{2fluss} differs considerably from the one in Fig. \ref{2fluspu}: even large size particles $a\sim 0.1-0.3~\mu$m  get heavily destroyed within $\sim (1-2)\tau_c(T=10^8)$. It appears that the required conditions for a dusty gas to shield its own dust particles against sputtering would be a strong interrelation between $\tau_{sp}$ and $\tau_s$. The relation $\tau_{sp}(a)/\tau_s\gg 1$ even for the smallest particles that are known to contribute considerably to the cooling, would inhibit dust destruction.

\subsubsection{{Dust survival regime before the Sedov-Taylor phase}}\label{2flust}

The foregoing discussions show that the possibility of survival of dust behind the RS critically depends on the initial thermal state of the gas. If at the initial time, i.e., immediately behind the shock $\tau_{sp}(a=0.003~\mu{\rm m})>{\rm ~ a ~few}~\tau_{c}$, the situation can change significantly. In this case, if the smallest dust particles survive during the first couple of $\tau_c$, the initial cooling function will provide rapid cooling and prevent destruction of the most efficient cooling agents -- the small dust grains. This would be a sufficient condition for the `self-shielding' due to dust cooling. The dotted red curve in Fig. \ref{sputoco} shows the $\tau_{sp}/\tau_c$ for the cooling function from a MRN-type dust (with $n(a)\propto a^{-3/5}$ in $a=0.03-0.3~\mu$m) versus gas temperature for a dust particles of radius $a=0.1~\mu$m. At $T\simeq 10^7$ K, the ratio $\tau_{sp}/\tau_c\simgt 1$ for all dust particles in this size range. One can, therefore, conclude that dust cooling regime can operate in those epochs of the reverse shock when the post-shock temperatures $T_{rs}<10^8$ K, which are expected before establishing the Sedov-Taylor stage (see Appendix \ref{brs}). 

Figure \ref{2fluss7} illustrates the evolution of the partial dust-to-gas mass fraction for particles of different sizes. The upper panel presents the case with the gas metallicity in the ejecta $\zeta_{d,0}=0.01$ as in the Milky Way ISM, and as in the models considered above in Sections \ref{isolatesph} and \ref{kolmog}. Note that $\zeta_{d,0}=y_{sn}/M_{ej}$, where $y_{sn}$ is the SN dust yield in $\msun$ as defined above. The initial metallicity of ejecta in the model presented in the lower panel is assumed to be three times that of the Milky Way value $\zeta_{d,0}=0.03$, which corresponds to the predicted SN dust yield $y_{sn}\simgt 0.6~\msun$   \citep{Todini2001,Bianchi2007,Cherchneff2009,Cherchneff2010,Sarangi2013,Sarangi2015,Marassi2015,Marassi2019}.  We find that although  the fraction initially decreases, it reaches an asymptotic value at longer time scales, depending on the grain size. E.g., for an initial value of dust-to-gas mass ratio $\zeta_{d,0}=0.03$ (lower panel), the survival fraction of grains of size larger than $a \simgt 0.05 \,\mu$m can be as large as $0.5$ at longer time scales.

\begin{figure} 
\center
\includegraphics[width=7.5cm,angle=00]{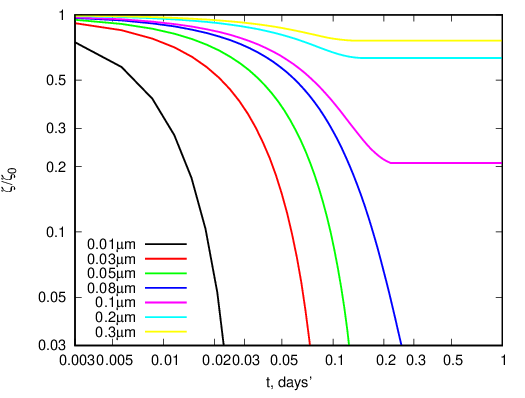}
\includegraphics[width=7.5cm,angle=00]{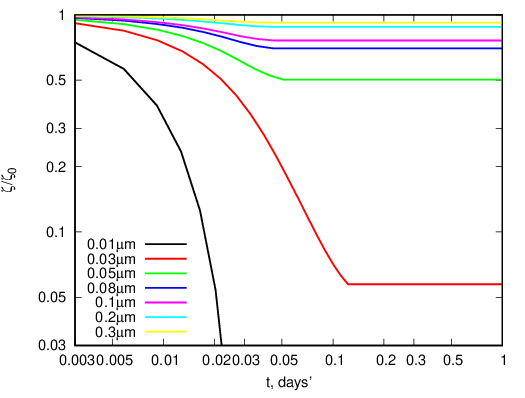}
\caption{Same as in Fig. \ref{2fluss} with the initial temperature $T=10^7$ K, and the initial dust-to-gas mass fraction $\zeta_{d,0}=0.01$ as in the MW in {\it {left} panel}, and $\zeta_{d,0}=0.03$ in the {\it {right} panel} for particle of different sizes as indicated in legends; in the verticle axis partial ratios $\zeta_{d}(a_i)/\zeta_{d,0}(a_i)$ are shown. 
}
\label{2fluss7}
\end{figure}

\begin{figure} 
\center
\includegraphics[width=7.5cm,angle=00]{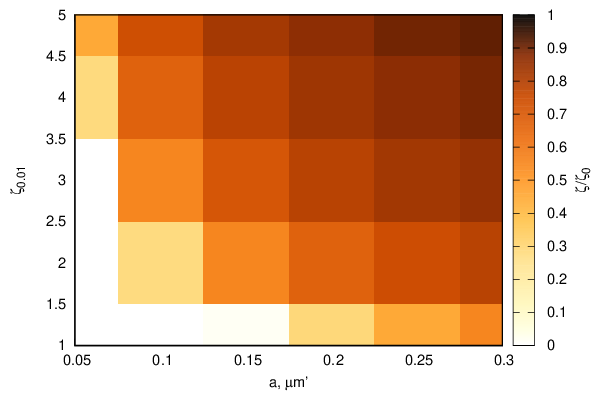}
\includegraphics[width=7.5cm,angle=00]{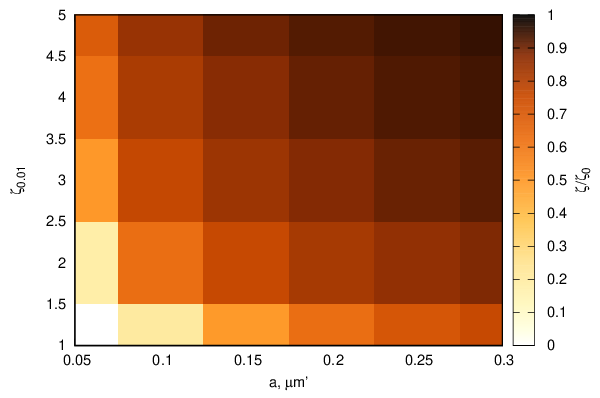}
\caption{Color bar in the right side shows the partial fraction of survivoing dust particles of different sizes indicated at $x$-axis $\zeta_d(a)$ to their initial values $\zeta_{d,0}(a)$ (before the RS), after processing by the RS with the post-shock temperature $T_{rs}=3\times 10^7$ K ({left} panel), and $T_{rs}=10^7$ K ({right} panel) . The $y$-axis shows the total initial dust-to-mass ratio $\zeta_d$ in the normalization coefficient $C$ in the cooling function (Eq. \ref{coolnorm}) in units of the Milky Way value $\zeta_{d,MW}=0.01$, as assumed.    
}
\label{2flucum}
\end{figure}

Figure \ref{2flucum} shows the dependence of the surviving dust-to-gas ratio of dust of a given radius to its initial value $\zeta_d(a)/\zeta_{d,0}(a)$ (shown with the help of a color bar on the right), as a function of the initial size (plotted in the $x$-axis). While the upper panel shows the case for $T_{rs}=3 \times 10^7$ K, the bottom panel is for $T_{rs}=10^7$ K. Note that the post-shock temperature in the models presented in Fig. \ref{2flucum} is assumed $T_{rs}=3\times 10^7$ K to lie intermediate between that in Fig. \ref{2fluss} and the one in Fig. \ref{2fluss7}.  The plot shows that (a) the survival fraction is larger for larger grains and larger initial value of $\zeta_d(a)$, and (b) the survival fraction is larger for the lower temperature case presented in the bottom panel than at $T_{rs}=3 \times 10^7$ K.

\begin{figure} 
\center
\includegraphics[width=6.0cm,angle=270]{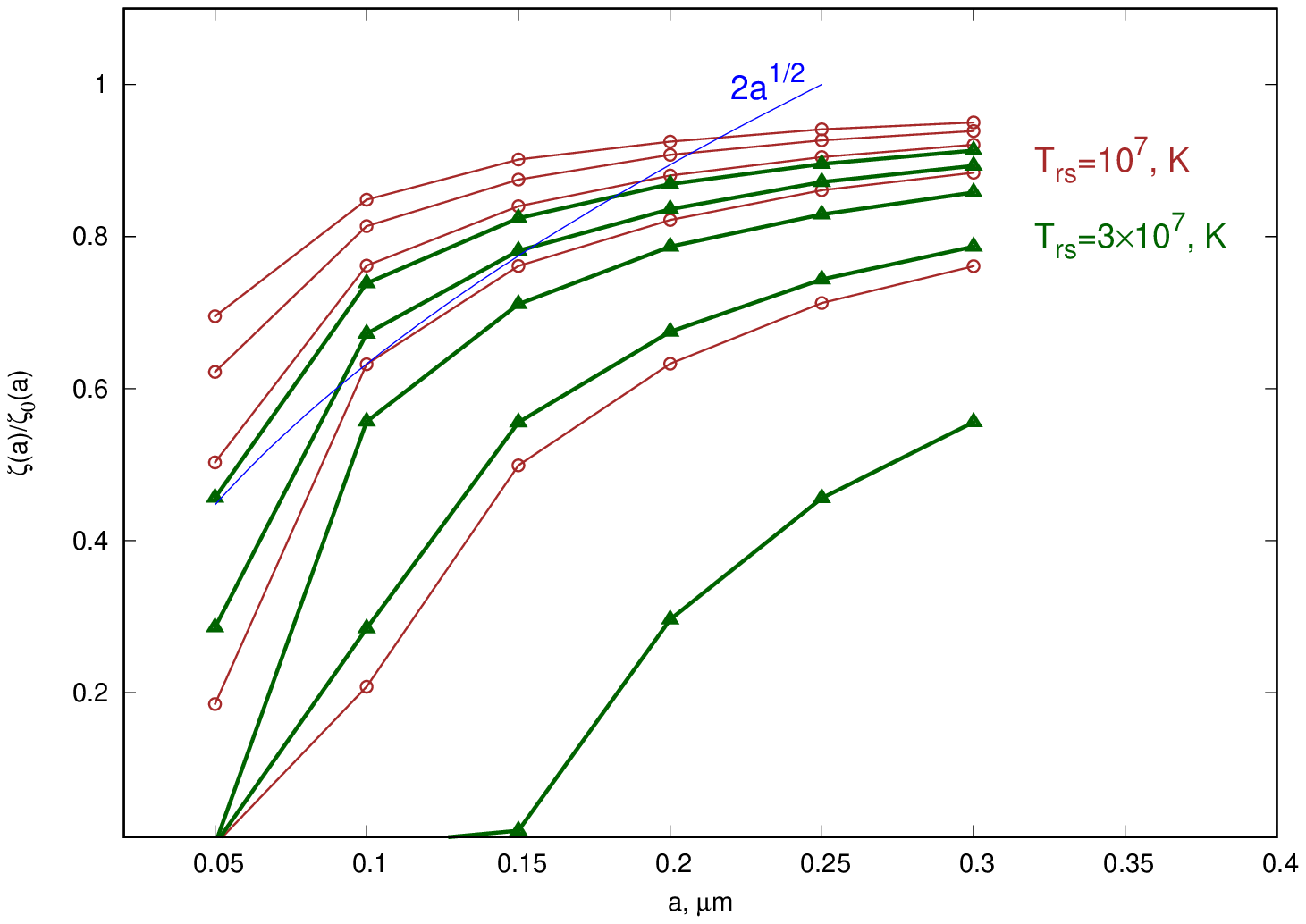}
\caption{Mass distribution of surviving dust particles behind reverse shocks of different properties $m(a)\propto a^3n(a)$,  after processing by the reverse shock. The $x$-axis represents the dust grain radius $a$, $y$-axis shows the mass fraction of surviving grains with a given radius $a$ relative to its initial value in the ejecta $\zeta(a)/\zeta_{d,0}(a)$. Thin brown curves represent dust particles processed by the reverse shock with  $T_{rs}=10^7$ K; the dark-green curves are for the  $T_{rs}=3\times 10^7$ K. From bottom to top, the curves of the two sets show the mass distribution $m(a)$ behind the RS, with the total dust-to-gas mass fraction $\zeta_d$ in the cooling function, see Eq. (\ref{coolnorm}): $\zeta_d=(1,~2,~3,~4,~5)\times \zeta_{d,MW}$ of the Milky Way value. Thin blue line $m(a)=2a^{1/2}$ corresponds to the size distribution $n(a)\propto a^{-2/5}$            
}
\label{2flumrn}
\end{figure}

Figure \ref{2flumrn} illustrates the dust mass distribution function $m(a)=4\pi a^3n(a)/3$ of dust particles that survive after being  processed by the reverse shocks of different post-shock temperatures. The initial distribution function is assumed to be an MRN type $m(a)\propto a^{-1/2}$. It is readily seen that the size distribution function of the surviving particles is  considerably flatter, from $n(a)\propto a^{-2}$ to $\propto a^{-2.5}$ for particles of $a<0.15~\mu$m, and $\propto a^{-3}$ for larger particles $a\geq 0.15~\mu$, as compared to a MRN distribution $n(a)\propto a^{-3.5}$. 

The most noteworthy feature of the post-RS dust processing is that the very possibility of dust survival against the sputtering from the hot gas is determined by the radiation cooling provided by the dust itself. As a consequence, the higher amount of dust nucleated before the impact from the RS produces favorable conditions for dust survival as seen in Fig. \ref{2flumrn}, while going from curves with a lower to  higher initial values of $\zeta_d$ in the cooling function. From this point of view, dust cooling tends to result in a `dichotomy' in the dust-to-stellar mass ratio being either `super-dusty' with $\zeta_\ast \sim 0.01$ for models with $\zeta_{d,0}\simgt 5\zeta_{d,MW}$, or dust-deficient galaxies with the observationally inferred value of $\zeta_\ast \ll 0.01$.  

\section{{Summary}}\label{summ}

Dust produced in supernovae ejecta at the end of the adiabatic expansion is commonly thought to be destroyed by thermal sputtering under the action of the reverse shock wave, which heats the ejecta up to $T\sim 10^8$ K. This scenario  conflicts with the observations of dust overabundance with the dust-to-stellar mass ratio $\zeta_\ast\simgt 10^{-2}$ in galaxies at the beginning of reionization epoch, $z\sim 8-9$. In order to mitigate this conflict we have considered the possibility of a SNe produced dust to survive against the sputtering under the influence of  radiative energy losses induced by an enhanced infrared emission from the dust already formed before being hit by the reverse shock. We have found that: 

\begin{itemize}
\item A predominance of the radiative cooling at $T\simgt 3\times 10^6$ K connected with an efficient conversion of thermal energy of the hot plasma into infrared radiation is  critically important for thermal evolution of the gas behind the reverse shock. The flat temperature dependence of the `dust cooling' function ($\Lambda(T)\approx$const) at $3\times 10^6 \leq T\leq 3\times 10^7$ K can induce rapid radiation cooling and  thermal instability with a characteristic cooling time shorter than the dust sputtering time for particles with radii $a\simgt 0.05~\mu$m. At early stages of the reverse shock before the Sedov-Taylor stage, when the gas temperature behind the RS is relatively low $T_{ps}\sim 10^7$ K, a considerable dust mass fraction (up to $\simgt 0.5$) of large particles with $a>0.05~\mu$m survives sputtering. 

\item However, hotter regions of the RS with $T_{rs}>3\times 10^7$ K destroy dust particles on shorter times scales than radiative cooling time scale, and considerably inhibit the dust cooling. This results in a more efficient, by factor $\sim 2$, dust destruction as compared to the previous case. Later stages with $T_{rs}\sim 10^8$ K are more hostile to the dust particles: only $\sim 30$\% of the largest particles survive at times longer than one cooling time.

\item  A larger amount of nucleated dust grains proportionally enhances the cooling function. This makes the hot post-RS gas to cool faster below the sputtering threshold. In other words, the higher the dust mass produced via nucleation, the faster the radiation cooling and the higher the survival dust fraction. Such a `positive feedback' between radiation cooling and dust survival makes the net dust SN yield very sensitive to the initial conditions in the ejecta.    

\
\end{itemize}

\section{Acknowledgements}
\label{acks}
We thank K. C. Sarkar, A. V. Tutukov and E. O. Vasiliev for friendly criticism and valuable comments. {We are thankful to F. Kirchschlager and F. D. Priestley for their kind criticism regarding dust in local young SNRs.} Continuous support in numerical simulations from E. O. Vasiliev is greatefully acknowledged. YS thanks the hospitality of the Raman Research Institute. Simulations have been performed with the RRI HPC cluster.


\appendix

\section{Conditions behind the RS}\label{brs}

The initial temperature and density in ejecta \citep{Truelove1999}

\be \label{teja}
T_{ej}\sim {2\over 3}{\mu m_{\rm H}E_B\over k_BM_{ej}}\simeq 5\times 10^8E_{51}M_{10}^{-1}~{\rm K}\,,
\ee 

\be \label{nej}
n_{ej}\simeq 7\times 10^{15}M_{10}R_{0,3}^{-3}~{\rm cm^{-3}}\,, 
\ee 
$M_{10}=M_{ej}/10~\msun$ is the ejecta mass, the initial radius of the ejecta is assumed to be $R_0=1000 R_\odot$, and $R_{0,3}=R_0/10^3~R_\odot$. The expansion velocity (Zeldovich \& Raizer) is 

\be \label{cej}
v_{exp}={2c_s\over {\gamma-1}}=\sqrt{15k_BT\over {\mu m_{\rm H}}}\,. 
\ee
We use $\gamma=5/3$, $\mu\simeq 0.5$. We also assume that dust particles begin to nucleate at $T_{nuc}\simeq 3000$ K \citep[][]{Todini2001}, which suggests the ejecta gas density to  be,

\be \label{nnuc}
n_{nuc}=n_0\left({T_{nuc}\over T_{ej}}\right)^{3/2}\simeq 10^{8} M_{10}^{5/2}R_{0,3}^{-3}E_{51}^{-3/2}~{\rm cm^{-3}}\,,
\ee
when its radius is  

\be \label{rnuc}
R_{nuc}\sim 3\times 10^{16}R_{0,3}E_{51}^{1/2}M_{10}^{-1/2}~{\rm cm}\,. 
\ee  

When the reverse shock forms and propagates inward, the gas temperature behind the RS is  
\be \label{tnuc}
T_{rs}\sim {3\over 16}{\mu m_{\rm H}\tilde v_{r}^2\over k_B}\,,  
\ee  
where $\tilde v_r\equiv v(R_{rs})-v_{r}$ is the velocity of the reverse shock in the rest frame of the expanding un-shocked ejecta \citep[see Fig. 1 in ][]{Mckee1995,Truelove1999}. In the uniform ambient medium, with  density $\rho_0=$ const, at the early stage of evolution -- before the Sedov-Tayor stage ( $t< t_{\rm ST}$), one has, 
\be 
\tilde v_r\simeq 5.4^{3/2}(1+2.95t^{3/2})^{-5/3}, 
\ee 
here $t$ and $\tilde v_r$ are given in the characteristic units $t_{ch}=3\times10^3 E_{51}^{-1/2} M_{10}^{5/2}n_0^{-1/3}$ yr, $v_{ch}\simeq E_{51}^{1/2}M_{10}^{-1/2}$ km s$^{-1}$. At the Sedov-Taylor stage $t\geq t_{\rm ST}$  
\be 
\tilde v_r\simeq 0.533+0.106t.
\ee 
The corresponding gas temperature behind the RS is 
\be 
T_{rs}\simlt 10^7, ~{\rm K,~ at}~t< t_{\rm ST}\,, 
\ee     
which later changes to,
\be 
T_{rs}\geq 7\times 10^7~{\rm K,~at}~t\geq t_{\rm ST}\,.
\ee

\section{Radiative-conductive interface between the ejecta and hot wind bubble}\label{interf} 

A fraction of hot particles from the wind bubble can penetrate into the denser ejecta, heat it and destroy dust  grains in it via thermal and kinetic sputtering. The mean free path of protons in the wind bubble is $\ell_p\sim 5\times 10^{19}T_8^2n_w^{-1}$ cm. Taking the value of the Coulomb logarithm  as $\ln\Lambda=25$, one has  $\ell_p\sim 3\times 10^{21}T_8^2n_0^{-1}$ cm, where $T_8=T/10^8$ K, and $n_w\sim 10^{-3}n_0$, is the density in the wind bubble. Even for an ambient density of $n_0\sim 10^3$, the free path length $\sim 1$ pc is much larger than the ejecta radius at the beginning of dust nucleation: $R_{nucl}\sim R_{ej}\sqrt{T_{ej}/T_{nucl}}\sim 7.5\times 10^{16}(E_{51}/M_{10})^{3/4}$, where we have used $T_{ej}\sim 8\times 10^8(E_{51}/M_{10})^{1/2}$ K, and $R_{ej}\sim 10^{14}$ cm, and $M_{10}=M_{ej}/10~\msun$. This implies a heat flux into the ejecta from the hot gas,  $q\sim \rho_wu_{th}^3\sim 2\times 10^{-3}n_0T_8^{3/2}$ erg cm$^{-2}$ s$^{-1}$.  

{At the time of nucleation , the gas temperature and density in the ejecta are $T=T_{nuc}\sim 10^3$ K and $n_{nuc}\sim 10^8M_{10}^{5/2}R_{0,3}^{-3}E_{51}^{-3/2}$ cm$^{-3}$ [Eq. (\ref{nnuc}) in \ref{brs}]. The mean free path  of hot protons entering into the ejecta drops by factor $\sim n_{nuc}/n_w\sim 10^8$. Protons can propagate inwards only diffusively in such conditions. As a result, the incoming heat $q$ is immediately  lost through radiation within the boundary layer of thickness, 
\be 
\Delta R\sim {2q\over \Lambda(T_{nuc})n_{nuc}^2}\sim 4\times 10^7~{\rm cm}\,,  
\ee
with the cooling function $\Lambda(T)\sim 10^{-26}$ erg cm$^{3}$ s$^{-1}$ at $T=T_{nuc}\sim 10^3$ K \citep[Fig. 30.1 in ][]{Draine2011}.
}

\section{Heating efficiency of a single dust particle}\label{hat}

Following \citet[][]{Dwek1987} the efficiency of heating $h(a,T)$ can be approximated as 

\begin{equation}
h(a,T)\simeq 
\begin{cases}
1\,, & T<T(a)\\
[T(a)/T]^{3/2}\,, & T>T(a)
\end{cases} 
\end{equation}
here $T(a)\simeq 2\times 10^5(a/0.0005~\mu{\rm m})^{3/4}$.

\section{Explicit solution of dust sputtering}\label{explic}

After the corrector step, we get the gas and dust densities $\rho_g^{i+1} = m_p n^{i+1}$, $\rho_d^{i+1}$ and the gas temperature $T^{i+1}$. The dust mass in a cell is  $M_d^i = \rho_d^{i+1} dV$, where $dV$ is the volume of a cell. The mass of each grain in a cell is $m_d^i = 4\pi \rho_m (a^i)^3 / 3$, where $\rho_m=2.2$ g/cm$^3$ is the grain material density, $a^i$ is the grain size at $i$-th step. The number of grains in a cell is $N_d = M_d^i / m_d$.

The equations for the mass and size of a grain are
\be
 m_d^{i+1} = m_d^i - 4 \pi \rho_m {\cal R}_d(n^{i+1},T^{i+1}) (a^i)^2 \Delta t^{i}\,,
\ee
\be
 a_d^{i+1} = a_d^i - {\cal R}_d(n^{i+1},T^{i+1}) \Delta t^{i}\,,
\ee
where ${\cal R}_d(n^{i+1},T^{i+1})$ is the destruction rate. Assuming conservation of the number of grains in time, the mass of dust in a cell is $M_d^{i+1} = m_d^{i+1} N_d$ and the dust density is $\rho_d^{i+1} = M_d^{i+1} / dV$.





\end{document}